\newcommand{\green}{f_{_{\rm G}}}
\documentstyle[psfig]{article}
\font\t=cmbx10 at 16pt
\font\au=cmr10
\font\ad=cmti10 at 9pt
\font\s=cmbx10

\font\fc=cmr10 at 9pt
\def\pmb#1{\setbox0=\hbox{$#1$}%
\kern-.025em\copy0\kern-\wd0
\kern.05em\copy0\kern-\wd0
\kern-.025em\raise.0433em\box0}

\begin{document}
\begin{center}
{\t Electron acceleration in solar noise storms}\\
\end{center}
                                                                                
\begin{center}
{\au  Prasad Subramanian}\\
{\ad Indian Institute of Astrophysics, Koramangala, Bangalore- 560034, India\\
e-mail: psubrama@iiap.res.in}
\end{center}
\vspace{.8cm}

{\au 
\noindent
{\s Abstract}
\\
\\
We present an up-to-date review of the physics of electron acceleration in solar noise storms. We describe the observed characteristics of noise storm emission, emphasizing recent advances in imaging observations. We briefly describe the general methodology of treating particle acceleration problems and apply it to the specific problem of electron acceleration in noise storms. We dwell on the issue of the efficiency of the overall noise storm emission process and outline open problems in this area.
\\
\\
\noindent 
{\s 1. Introduction:} 
\\
\\  
\noindent 
{\s 1.1 Motivation:}
\\
\\
\noindent
Noise storms are the most common form of meter wavelength radio emission from the solar corona. The nomenclature arises from hissing sounds produced in short-wave radio receivers, and was coined around the 1930s. Noise storms are sites of long-lasting quasi-continuous electron acceleration in the solar corona, and we will focus on this aspect here. Electron acceleration (and particle acceleration in general) is of central importance in several astrophysical problems. A thorough understanding of this process in the solar corona can therefore be of considerable use in understanding its import in objects that are farther away and less accessible to observations.
\\
\\
\noindent 
{\s 1.2 Brief History:}
\\
\\ 
\noindent
The recognition that the sun could be a source of intense meter wavelength emission took place around 1942, when the operations of British anti-aircraft radar were severely affected by such emissions. However, owing to the then-ongoing war, the scientific results were not published until 1946 (Hey[4]), and this heralded the birth of the rich field of solar radiophysics. A good overview of meter wavelength solar phenomena can be found in McLean \& Labrum[5]. More recent overviews of solar radio emission that are not confined only to meter wavelength phenomena can be found in Gary \& Keller[3] and Bastian \& Gary[1].
\\
\\
\noindent 
{\s 1.3 Brief overview of solar meter wavelength emission}
\\
\\
\noindent
There are several kinds of emission from the solar corona at meter wavelengths, which have distinct observational signatures. Some examples are given in figures 1 and 2. 
\begin{figure}[h]
\centerline{\psfig{figure=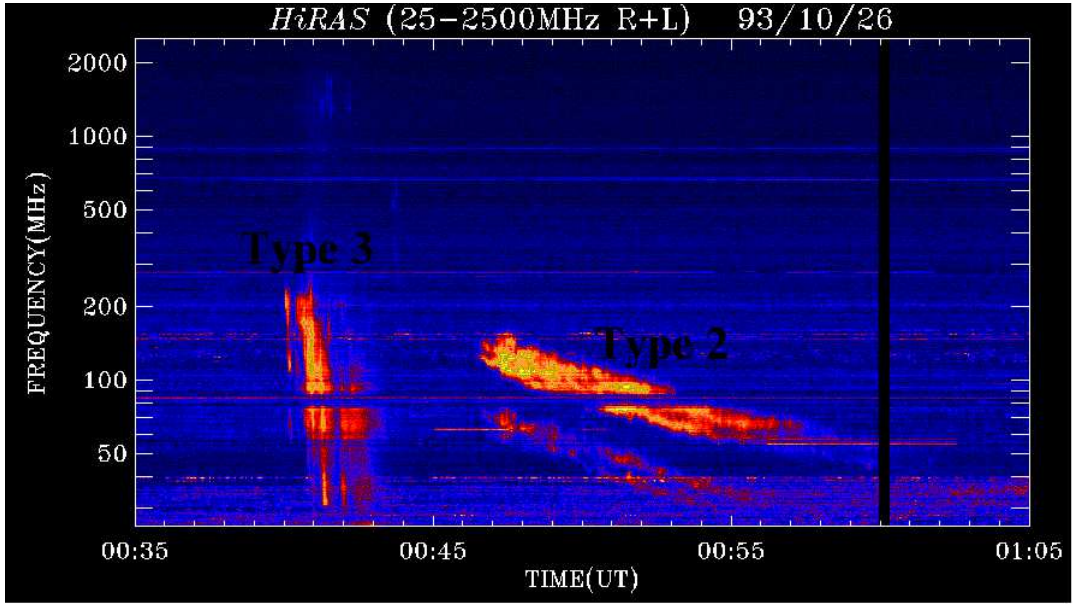,height=8cm}}
\caption{\fc Examples of types 2 and 3 meter wavelength radio emission: This is a dynamic spectrum from the Hiraiso radio spectrograph in Japan. Time runs along the $x$-axis, and frequency along the $y$-axis. The intensity from the whole sun is expressed by the colorscale.}
\end{figure}
\begin{figure}[h]
\centerline{\psfig{figure=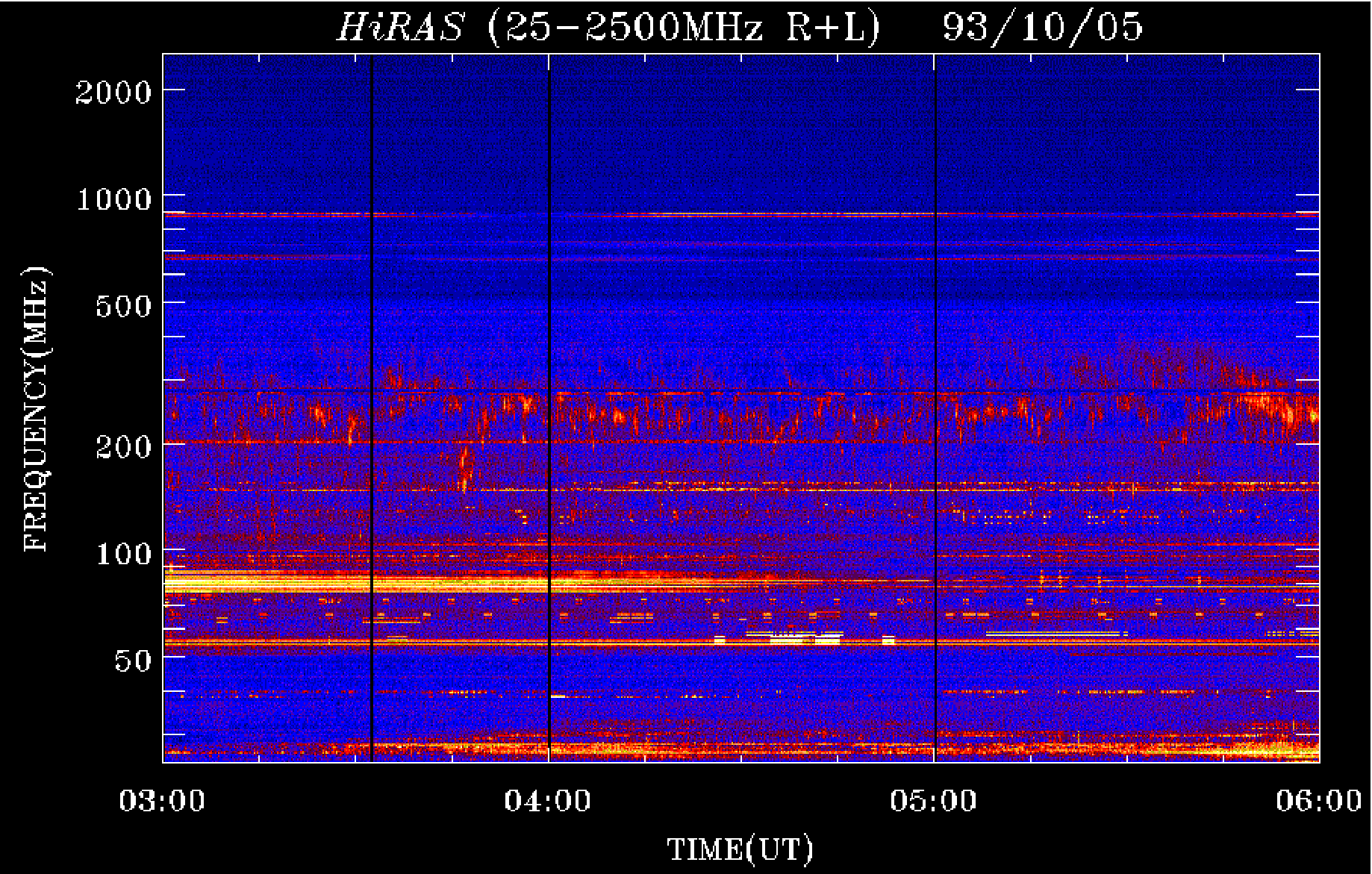,height=8cm}}
\caption{\fc Example of type 1 emission: rest of the caption same as that for figure 1.}
\end{figure}
These figures are called {\em dynamic spectra}. They are multifrequency records of intensity from the entire sun, and contain no spatial information. Time runs along the $x$-axis of these figures and the observing frequency is along the $y$-axis. The observed intensity is represented by a colorscale/grayscale. Different kinds of emission are characterized by different kinds of characteristic signatures on such dynamic spectra. In figure 1 for instance, the emission labelled 'type 2' has a characteristic signature where the bright emission drifts downwards in frequency with time. It is taken to be a signature of electrons accelerated at a shock front that is travelling outwards through the solar corona. The kind of emission labelled 'type 3' comprises of bright, almost vertical tracks on the dynamic spectrum; this kind of emission is taken to be a signature of relativistic electrons escaping outwards through the solar corona along open magnetic field lines. On the other hand, the kind of emission depicted in figure 2 is called noise storm (or type 1) emission. It is rather unspectacular, and comprises of a broadband {\em continuum} that lasts for several hours to days on which there are superposed several randomly distributed, short timescale (0.1--1 second) narrowband, intense {\em bursts}. This kind of emission is thought to be caused by nonthermal/accelerated electrons. We will concentrate here only on these noise storms. The reasons are twofold; on the one hand, they are the most common signature of accelerated electrons at meter wavelengths, and have been very well observed. Owing to their ubiquitous and long-lasting nature, they are also well suited to repeated observations. On the other hand, some important aspects of noise storm emission (both with regard to the continuum and bursts) that are still rather poorly understood, despite decades of study. Furthermore, the plasma emission process, (which we will describe later) which is thought to be operational in noise storm emission, is also central to other kinds of meter wavelength emission such as type 2, 3 and 5 emission. A thorough understanding of the noise storm phenomenon is therefore of considerable utility.
\\
\\
\noindent 
{\s 1.4 Organization of paper:}
\\
\\
\noindent
The rest of this paper is organized as follows: we describe further details of the observed characteristics of noise storm emission in \S~2. We pay particular attention to multifrequency observations of noise storms and to recent imaging observations that have the potential to significantly impinge on some long-unsolved theoretical issues. We next turn our attention to the physics of the emission process via which the observed noise storm radiation is produced in \S~3. We emphasize the role of accelerated electrons which are the starting point of the overall emission process. The main focus of this paper is the manner in which electrons are accelerated in a quasi-continuous manner to power noise storm emission at meter wavelengths. Accordingly, we review the development of standard theoretical treatments of particle acceleration in \S~4. We first elaborate on the physical scenarios where particles are accelerated, such as reconnection regions and shocks. We then trace the development of particle transport equations that treat diffusion in velocity/momentum space and yield the Fermi acceleration scenario. We then turn to the specific problem of electron acceleration in noise storm sources in the solar corona in \S~5. We estimate the power input to the accelerated electrons and compare it with the power observed in the observed radiation, deriving an efficiency for the overall process. In \S~6 we outline the applicability of such an efficiency estimate to other kinds of radio emission from the solar corona, and its role in furthering our understanding of solar coronal transients. We also include a brief discussion of open problems in this interesting area.

\vspace{.4cm}
\noindent
{\s 2. Observations of noise storm emission}
\\
\\
\noindent 
{\s 2.1 Multifrequency observations:}
\\
\\
\noindent 
The most basic identification of type1/noise storm emission, as mentioned earlier, is from dynamic spectra (figure 1). Type 1 emission is characterized by a long-lived (few hours to days), wideband ($\delta f/f \sim 100$\%) continuum, together with several intense, randomly interspersed short-lived (0.1--1s), narrowband ($\delta f/f \ll 1$) bursts. Early reviews on noise storms can be found in Wild et al.[8], Kundu[9] and Kruger[10]. An extensive review of noise storm observations is given by Elgaroy[2]. Noise storms are typically observed from 50 -- 500 MHz, and are brightest around 100-200 MHz. The detailed multifrequency characteristics of noise storms have not been studied very well; to the best of our knowledge, the only such study after the early pioneering study of Smerd[7] have been those of Thejappa \& Kundu[13], Kundu \& Gopalswamy[12] and Sundaram \& Subramanian[6]. While Smerd's[7] early results showed that noise storms tended to be brightest around 100 MHz, with intensities tapering off on either side of this (approximate) frequency, Sundaram \& Subramanian's[6] more recent, relatively sophisticated studies were confined only to the 50--80 MHz frequency range, and showed that noise storm intensities clearly rose as a function of frequency in this range. 
We will have occasion to comment on the interesting implications of such multifrequency observations in \S~6.
\\
\\
\noindent 
{\s 2.2 Imaging observations of noise storms:}
\\
\\
\noindent
Early imaging observations of noise storms were made with the Culgoora radioheliograph in Australia (e.g., Dulk \& Nelson[11]). Later on, noise storms were imaged with the Clark Lake Radioheliograph in the USA (e.g., Kundu \& Gopalswamy[12]; Thejappa \& Kundu[13]). Around the same time, the Nancay radioheliograph (NRH) in France also carried out extensive noise storm observations (e.g., Kerdraon \& Mercier[14]; Malik \& Mercier[15]). Occasional noise storm observations are also carried out with the Very Large Array (VLA) in the USA (Willson, Kile \& Rothberg[16]; Willson[17]; Habbal, Ellman \& Gonzalez[18]; Habbal et al.[19]). Lately, there have been attempts at combining visibilities from the Giant Metrewave Radio Telescope (GMRT) in Pune, India with those from the NRH to obtain images of noise storm sources in the solar corona. This approach combines several complementary advantages offered by the solar-dedicated NRH and the GRMT. It yields the highest dynamic range images of the solar corona at meter wavelengths (Mercier et al.[20]). Figures 3 and 4 show a couple of examples of results obtained by using this technique.
\begin{figure*}[h]
\centerline{\psfig{figure=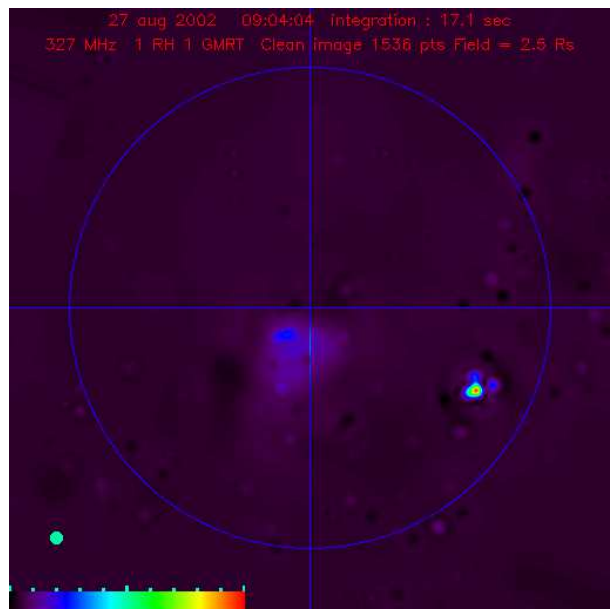,height=8cm}}
\caption{\fc 17 second snapshot of the solar corona at 327 MHz around 09:04 UT on Aug 27 2002. The resolution of this image is 49$^{''}$ and the rms dynamic range is 283. The bright noise storm emission in the southwestern quadrant is clearly evident, as are the two weak sources and the intervening diffuse emission near disk center. This image was made by combining visibilities from the NRH and the GMRT (Mercier et al.[20]).}
\end{figure*}
\begin{figure*}[h]
\centerline{\psfig{figure=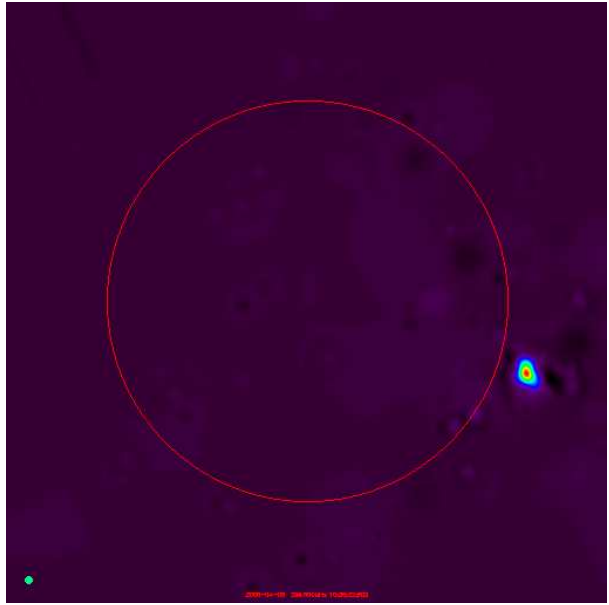,height=8cm}}
\caption{\fc 2 second snapshot of the solar corona at 236 MHz around 10:36 UT on Apr 06 2006. The resolution is 29$^{''}$ and the rms dynamic range 447. This image was made by combining visibilities from the NRH and the GMRT (Mercier et al.[20]).}
\end{figure*}
This technique allows dynamic ranges of 300-500 to be achieved with integration times of a few seconds. Such dynamic ranges were previously obtained only with integration times of around 3--4 hours. High dynamic ranges allow simultaneous imaging of bright and dim features, as is evident from figure 3. This feature is rather important in several situations, for bright noise storms are often accompanied by relatively dim, large-scale features such as coronal mass ejections (e.g., Bastian et al.[21]; Habbal et al.[19]; Willson[22]; Willson[23]). The high resolution afforded by this technique can be important in resolving the important question of angular broadening due to coronal turbulence (e.g., Bastian[24] and references therein). While there are already some hints that theoretical estimates of angular broadening might be somewhat exaggerated (Zlobec et al.[25]), this technique has the potential to set a definitive lower limit on the smallest observable source size in the solar corona. It can also set a firm upper limit on the brightness temperature of noise storms, which can have important implications for the plasma physics involved in the emission process (Robinson[26]; Kerdraon[27]).

\vspace{.4cm}
\noindent
{\s 3. Physics of noise storm emission process: overview}
\\
\\
\noindent 
As mentioned earlier, noise storms are sites of long-lasting, quasi-continuous electron acceleration in the non-flaring corona (e.g., Raulin \& Klein[29]; Klein[28] and references therein). They have been observed to occur in conjunction with other transient events such as coronal mass ejections and soft X-ray brightenings. In general, noise storms seem to be associated with emergence of new material in the corona and/or magnetic restructuring of some kind (e.g., Kerdraon et al.[30]; Bentley et al.[31]). Such rearrangements of magnetic fields are envisaged to take place via the process of magnetic reconnection (e.g., Priest \& Forbes[32]; Biskamp[33]), which results in the release of magnetic energy and consequent heating/acceleration of particles. Magnetic reconnection is also thought to responsible for acceleration of particles in violent eruptive events in the solar corona such as flares. Emerging magnetic flux into the solar corona can also drive an ensemble of weak shocks, which can accelerate electrons (Spicer, Benz \& Huba[34]). We will return to the subject of electron acceleration in the next section.

For now, we note that most of theoretical treatments of noise storm emission simply {\em assume} the presence of nonthermal electrons and proceed from there. If this nonthermal electron distribution has an anisotropy in velocity and/or physical space, it will emit an intense population of Langmuir waves through a mechanism analogous to Cherenkov emission (e.g., Robinson[26]; Mikhailovskii[35]; Melrose[36,37,38] and references therein). The anisotropy can be viewed as a source of free energy that is given to the population of Langmuir waves. The Langmuir waves (also referred to as plasma waves) comprise bodily oscillations of the plasma. These waves coalesce with a suitable population of low frequency waves such as ion-sound waves or lower hybrid waves (Wentzel[39]) in order to produce the observed electromagnetic (radio frequency) emission. This overall picture is usually called the ``plasma emission'' hypothesis, and is usually invoked for high brightness temperature emission that necessitates a coherent emission mechanism.

It may be noted that all the theoretical requirements, starting from that of electron acceleration and including the plasma emission process, are much more severe for the very intense, short-lived (0.1--1 sec) type 1 bursts than they are for the long-lived (few hours to days) background continuum. Detailed measurements (Malik \& Mercier[15]; Krucker et al.[40]) show that the bursts are much more compact than the continuum source, and that they tend to move randomly within the continuum source. Type 1 bursts could be manifestations of ``nanoflares'', which are small elemental releases of energy via small-scale reconnections and are suspected to be responsible for heating the ambient solar corona (Mercier \& Trottet[41]). The physical origin of such bursts could be statistical nonequilibrium fluctuations in the coronal plasma, and there are few theories that attempt to treat the continuum and bursts in a self-consistent manner (e.g., Thejappa[42]).

\vspace{.4cm}
\noindent
{\s 4. Particle acceleration:}
\\
\\
\noindent 
{\s 4.1 Physical situations for particle acceleration: reconnection sites and shocks}
\\
\\
\noindent 
While there are several interesting questions to be answered with regard to noise storm emission, we will focus here only on the aspect of electron acceleration, which is the starting point for the entire process. We first briefly describe the physical situations where particles are typically accelerated and follow it up with typical theoretical treatments. It may be emphasized that particle acceleration in general is ubiquitous in astrophysics, and is responsible for several interesting phenomena, ranging from ultra-high energy cosmic rays to radio and high energy emission from extragalactic jets to several situations in the solar corona and the earth's bowshock and magnetotail. Even in the solar corona, particle acceleration is responsible for observations associated with diverse phenomena such as nonthermal radio bursts, flares and coronal mass ejections (see, for e.g., Aschwanden[43]).
In the solar corona, particles are thought to be accelerated primarily due to magnetic reconnection (Priest \& Forbes[32]; Biskamp[33]) and at shocks (e.g., O'C Drury[44]; Quenby \& Meli[45]; Malkov \& O'C Drury[46]; Jones \& Ellison[47]).

Reconnection can be broadly thought of as a process by which stressed magnetic fields rearrange their topology in order to relax to a lower energy configuration. The excess energy is partially expended in accelerating particles. From a microscopic standpoint, the process of reconnection involves two oppositely directed magnetic field lines coming very close to each other and eventually changing their connectivities. When oppositely directed magnetic fields come close to each other, there will be a large magnetic field gradient in the reconnection region, necessitating the presence of a strong electric field in the plane perpendicular to the one which contains the magnetic fields. This electric field is capable of accelerating particles (e.g., Onofri, Isliker \& Vlahos[48]; Turkmani et al.[49]; Arzner et al.[50]). However, this is a complex problem involving treatments in both the magnetohydrodynamic and kinetic regimes, and it is probably fair to say that we are only beginning to develop an understanding of this process. Direct electric field acceleration apart, there can also be substantial turbulence in the vicinity of the reconnection regions; specifically, the reconnection outflows are typically turbulent. Particles can resonate with part of the turbulent wave spectrum and gain energy via wave-particle interactions.

The solar corona is forever in a state of flux, and magnetic flux is constantly being added/removed or being moved around. There is thus plenty of scope for reconnection to occur and reorganize fields on several scales, both small and large. As such, reconnection is held responsible for several transient phenomena in the solar corona such as the initiation of flares and coronal mass ejections. Before we leave this discussion, it may be noted that the wide scope of magnetic reconnection and its crucial importance has spurred laboratory studies to investigate this phenomenon in detail (e.g., Yamada et al.[51]).

The other agent that is typically invoked for particle acceleration is a shock. Roughly speaking, a shock is formed when a pressure/temperature/density disturbance propagates through a medium at a speed that is greater than the characteristic speed for the propagation of small pressure disturbances in that medium. For an unmagnetized medium this means that shocks are typically formed when a disturbance travels at supersonic speeds. For magnetized media, the situation is somewhat more complicated, since there are several characteristic speeds to consider: the Alfven speed and the slow and fast magnetosonic speeds, for instance. The shocking agent can be a localized energy release, such as that in a flare, which causes what is referred to as a ``blast wave'' shock, akin to that in a supernova explosion. It can also be a piston, like a coronal mass ejection. The shock itself can be viewed as a propagating discontinuity (in all physical quantities such as temperature, density and pressure). There is usually some form of turbulence present at the shock front which enables particles to diffuse back and forth across it. A particle that diffuses from the upstream side of the shock onto the opposite (i.e., downstream) side will collide with scattering centers moving with the shock and gain energy. If it manages to diffuse back upstream and then back downstream, it will gain more energy in a second collision. This is a rough description of what is usually referred to as diffusive shock acceleration (e.g., O'C Drury[44]; Quenby \& Meli[45]; Malkov \& O'C Drury[46]; Jones \& Ellison[47]). 
As with reconnection, shock acceleration is a mechanism that is applicable for a wide variety of astrophysical phenomena, not to mention several observational features in connection with the solar corona. In the context of noise storms, electrons are presumed to accelerated by a series of weak shocks driven by the emergence of new magnetic flux into the corona from beneath the photosphere (Spicer, Benz \& Huba[34]). Emerging magnetic flux can also cause repeated episodes of small-scale reconnection, which can be accompanied by electron acceleration.
\\
\\
\noindent 
{\s 4.2 Mathematical treatment of particle acceleration:}
\\
\\
\noindent
Having discussed the physical situations where particles can be accelerated, we now turn our attention to theoretical methods of treating this process. We start with the collisionless Boltzmann equation
\begin{equation}
\frac{D f}{D t} = \frac{\partial f}{\partial t} + \frac{\partial x}{\partial t}\frac{\partial f}{\partial x} + \frac{\partial p}{\partial t}\frac{\partial f}{\partial p} = 0\, .
\label{eq1}
\end{equation}
The quantity $x$ represents a spatial coordinate and $p$ represents a momentum coordinate and the distribution function $f$ is normalized as follows to yield the number density $n$:
\begin{equation}
n = \int\,dp\,f(p)
\label{eq2}
\end{equation}
For the sake of simplicity, we have written the Boltzmann equation with only one space and momentum dimension; it can easily be generalized to include three dimensions in each of these quantities. The intepretation of equation~(\ref{eq1}) is simple: it says that, in the absence of collisions, the number of particles in the $x$-$p$ phase space is conserved. In other words, the shape of the elemental volume $dx\,dp$ itself can be distorted with time, but as long as there are no collisions, the number of particles in this volume will stay constant. It is worth noting that the well-known thermal Maxwellian distribution is an equilibrium solution of the collisionless Boltzmann equation. In other words, any initial particle distribution will eventually relax to a Maxwellian, given enough time. However, this is a highly idealized equation; in reality, there will be several kinds of collisions that can move particles in and out of the $dx\,dp$ volume. These can be represented on the right hand side of the Botlzmann equation as
\begin{equation}
\frac{D f}{D t} = - C_{\rm out} + C_{\rm in} = \frac{\partial f}{\partial t}\,\bigg |_{\rm c}\, ,
\label{eq3}
\end{equation}
where $C_{\rm in}$ represents the flux of particles entering and $C_{\rm out}$ the flux of particles leaving the phase space as a result of collisions. It may be noted that these collisions need not actually be physical collisions; they can be any process that scatter particles in an elastic or inelastic manner. For instance, they could represent intermittent episodes of acceleration that a particle might experience in moving through reconnection regions or interactions with part of a turbulent wave spectrum with which the particle resonates. If the cumulative effect of several small deflection collisions dominates over large deflection collisions, it turns out that the collision term (eq.~\ref{eq3}) can be conveniently written in what is called a {\em Fokker-Planck} form. A rigorous derivation of the Fokker-Planck equation can be found in plasma physics textbooks such as Sturrock[52] and Montgomery \& Tidman[53]. The Fokker-Planck equation can also be heuristically motivated by considering a situation where stochastic forces acting on a typical particle cause its momentum vector to execute a random walk as a function of time and diffuse in momentum space. In addition to this, the particle would also experience a drag/frictional force while moving through a dilute medium. The Fokker-Planck equation for such a situation can then be written as
\begin{equation}
\frac{\partial f}{\partial t}\,\bigg |_{\rm c} = - \frac{\partial}{\partial p}\biggl ( B f \biggr ) + \frac{1}{2} \frac{\partial^{2}}{\partial p^{2}}\biggl ( C f \biggr ) \, ,
\label{eq4}
\end{equation}
where the first term on the right hand side is the frictional term and the second represents diffusion in momentum space. The solution of equation~(\ref{eq4}) with only the first term of the right hand side would be a Gaussian in momentum space whose mean keeps decreasing with time; hence the nomenclature 'drag' term. On the other hand, if only the second term on the right hand side were retained, the solution would be a Gaussian whose variance (in momentum space) increases as a function of time, while its amplitude decreases so as to conserve the area under the curve. This shows why it is called the diffusion term. Although the frictional term as written in equation~(\ref{eq4}) represents energy loss, it could well represent energy gain if $B$ is negative; this represents a process called {\em first order Fermi acceleration}, and occurs in situations where the scattering centers move systematically, imparting energy to the particle. The import of the second (diffusion) term on the right hand side of equation~(\ref{eq4}) is somewhat less obvious; however, if it is recognized that the average momentum of a particle is the first moment of the distribution function (the zeroth moment is the number density, as per eq~\ref{eq2}), it can be understood that particles diffusing towards higher momenta are preferentially weighted, so that the mean momentum actually increases (see appendix of Subramanian, Becker \& Kazanas[54]). This effect is referred to as {\em second order Fermi acceleration}. Fermi originally conceived these acceleration processes (especially the second order process) from a kinematic point of view, as a consequence of a test particle bouncing off stochastically moving clouds. A treatment of Fermi acceleration from this viewpoint can be found in Longair[55]. In addition to these effects, there can also be particles escaping from and also injected into the system. The full transport equation including these terms can be written as (Becker, Le \& Dermer[56]):
\begin{equation}
\frac{\partial f}{\partial t}\,\bigg |_{\rm c} = - \frac{1}{p^{2}}\frac{\partial}{\partial p} \biggl ( p^{2} \biggl [ A(p)f - D(p)\frac{\partial f}{\partial p} \biggr ] \biggr ) - \frac{f}{t_{\rm esc}(p)} + S(p,t) \, ,
\label{eq5}
\end{equation}
where the first term on the right hand side is and equivalent representation of the right hand side of equation~(\ref{eq4}). The second term on the right hand side represents particle escape from the system with a mean timescale $t_{\rm esc}$ while the last term on the right hand side represents particle injection into the system.

This completes our cursory overview of the mathematical formulation of particle acceleration. In addition to what we have described, it may be mentioned that there can be additional terms on the right hand side of equation~{\ref{eq5}), such as diffusion in physical space and losses of various kinds (e.g., Becker[57]). There are a variety of numerical and analytical techniques that are employed to solve such particle transport equations. Much effort is devoted to calculating various parameters of the equation such as the diffusion coefficient in momentum space, escape timescale, etc., for a given physical situation. For an extensive treatment of such transport equations and various methods of solution, see Schlickeiser[58].

\vspace{.4cm}
\noindent
{\s 5. Electron acceleration in noise storms:}
\\
\\
We now focus on the specific problem of electron acceleration in solar corona. We confine our attention primarily to type I noise storm
{\em continua}, rather than the sporadic type I bursts, because we are
interested in examining the basic energetics of the electron
acceleration processes responsible for producing the quasi-continuous
radio emission. Most theories of type I phenomena invoke nonthermal electrons as a
crucial ingredient in producing the observed radiation. However, little attention has been focused on this problem in the previous literature. The majority of the theories
simply assume that nonthermal electrons are present, and focus most of
their attention on examining the wave-wave interaction processes through
which observable radio emission is ultimately produced. Although there is currently no theoretical consensus regarding the
fundamental mechanism powering type I phenomena, there is no question
that nonthermal electrons are responsible for the observed emission. In
view of the considerable uncertainty surrounding the precise physical
mechanism that results in the formation of the nonthermal portion of the
electron distribution, we follow the work of Subramanian \& Becker[59,60] in characterizing the acceleration in terms of a generic, stochastic
(second-order) Fermi process such as the one described in \S~4.
We first note that the total electron number density, $n_e$,
including both thermal and nonthermal particles, is related to the momentum
distribution $f$ via
\begin{equation}
n_e \, ({\rm cm^{-3}}) \, = \, \int_0^\infty p^2 \, f \, dp \ ,
\label{eq6}
\end{equation}
where $p$ is the electron momentum. This is somewhat different from the normalization expressed in equation~(\ref{eq2}). The associated total electron energy
density is given by
\begin{equation}
U_e \, ({\rm erg\,cm^{-3}}) \, = \, \int_0^\infty \epsilon \, p^2
\, f \, dp = \frac{1}{2 m_e} \, \int_0^\infty \, p^4 \, f \, dp \ ,
\label{eq7}
\end{equation}
where $m_e$ is the electron mass and $\epsilon = p^2/(2 m_e)$ is
the electron kinetic energy. We are mainly
interested in the nonthermal electrons, which are picked up from the
thermal population and subsequently accelerated to high energies.
In the situation where the acceleration timescale is smaller than any relevant loss timescales, 
the high-energy tail of the
electron distribution function is governed by a rather simple transport
equation that describes the diffusion of electrons in momentum space due
to collisions with magnetic scattering centers. The validity of this assumption is examined in detail in Subramanian \& Becker[59]. The time evolution of
the Green's function for this process, $\green$, is described by
\begin{equation}
{\partial \green \over \partial t} =
\frac{1}{p^2}\,\frac{\partial}{\partial p}\left( p^2 \, {\cal D}\,
\frac{\partial \green}{\partial p}\right) + {\dot N_0 \, \delta(p-p_0)
\over p_0^2} - {\green \over \tau} \ ,
\label{eq8}
\end{equation}
where ${\cal D}$ is the (as yet unspecified) diffusion coefficient in
momentum space and $\tau$ is the mean residence time for electrons in
the acceleration region. The source term in equation~(\ref{eq8})
corresponds to the injection into the acceleration region of $\dot N_0$
particles per unit volume per unit time, each with momentum $p_0$.
Although we do not explicitly include losses due to the emission of
Langmuir/upper hybrid waves by the accelerated electrons, it is expected
that these waves will be generated as a natural consequence of the
spatial anisotropy of the electron distribution (e.g., Thejappa[42]).
Note that in writing equation~(\ref{eq8}), we have ignored spatial
diffusion so as to avoid unnecessary mathematical complexity.

Although the specific form for ${\cal D}$ as a function of $p$ depends
on the spectrum of the turbulent waves that accelerates the electrons
(Smith[61]), it is possible to make some fairly broad generalizations
that help to simplify the analysis. In particular, we point out that a
number of authors have independently suggested that ${\cal D} \propto
p^2$. Examples include the treatment of particle acceleration by
large-scale compressible magnetohydrodynamical (MHD) turbulence (Ptuskin[62]; Chandran \& Maron[63]); analysis of the acceleration of electrons
by cascading fast-mode waves in flares (Miller, LaRosa \& Moore[64]);
and the energization of electrons due to lower hybrid turbulence (Luo, Wei \& Feng[65]). Hence we shall write
\begin{equation}
{\cal D} = D_0 \, p^2 \, ,
\label{eq9}
\end{equation}
where $D_0$ is a constant with the units of inverse time.
The resulting Green's function for the nonthermal electron
distribution is found to be (Subramanian \& Becker[59]; Subramanian, Becker
\& Kazanas[54])
\begin{equation}
\green(p,p_0) = A_0 \, \cases{
(p/p_0)^{\alpha_1} \ , & $p \le p_0$ \ , \cr
\phantom{space} \cr
(p/p_0)^{\alpha_2} \ , & $p \ge p_0$ \ , \cr 
}
\label{neweq2}
\end{equation}
where $p$ is the electron momentum, $p_0$ is the momentum of the
injected mono-energetic electrons, and the exponents $\alpha_1$ and
$\alpha_2$ are given by
\begin{equation}
\alpha_1 \equiv - {3 \over 2} + \left({9 \over 4}
+ {1 \over D_0 \, \tau}\right)^{1/2} \ , \ \ \ \ \
\alpha_2 \equiv - {3 \over 2} - \left({9 \over 4}
+ {1 \over D_0 \, \tau} \right)^{1/2} \ .
\label{neweq3}
\end{equation}
The quantity $\tau$ in these expressions represents the mean residence
time for electrons in the acceleration region, and the normalization
parameter $A_0$ is computed using
\begin{equation}
A_0 \equiv {\dot N_0 \over 2 \, D_0 \, p_0^3} \left({9 \over 4} +
{1 \over D_0 \, \tau}\right)^{-1/2} \ ,
\label{neweq4}
\end{equation}
where the constant $\dot N_0$ denotes the number of electrons injected
per unit time per unit volume into the acceleration region. The value of
the total electron number density associated with the Green's function
distribution is given by
\begin{equation}
n_{_{\rm G}} \equiv \int_0^\infty p^2 \, \green  \, dp = \dot N_0 \, \tau \ ,
\label{neweq5}
\end{equation}
as expected in this steady-state situation. Since the nonthermal
electrons are nonrelativistic with kinetic energy $p^2/(2 \, m_e)$, it
follows that we must require $\alpha_2 < -5$ in order to avoid an
infinite energy density, and therefore $D_0 \, \tau < 10^{-1}$.
Subramanian \& Becker[59] applied the formalism described above to model the
transport in momentum space of electrons injected from the high-energy
portion of the Maxwellian distribution in the corona. They found that
stochastic acceleration of the electrons dominates over losses due
to collisions and Langmuir damping for particles with momenta
$p > p_c$, where
\begin{equation}
p_c \equiv 5.35 \times 10^{-22} \left(\Lambda \, n_e \over D_0
\right)^{1/3}
\label{neweq6}
\end{equation}
denotes the ``critical momentum'' in cgs units, $\Lambda$ is the Coulomb
logarithm, and $n_e$ represents the total electron number density in the
corona. Electrons with $p > p_c$ experience net acceleration on average,
and those with $p < p_c$ are decelerated on average. It is expected that
a ``gap'' distribution will form as a result of the collisional and
Langmuir losses experienced by electrons with $p < p_c$.

Based on analysis of a generic second-order Fermi (stochastic)
acceleration mechanism, Subramanian \& Becker[59] demonstrated that if all of the
electrons in the Maxwellian distribution with $p > p_c$ are subject to
acceleration, then the nonthermal electron fraction is given by
\begin{eqnarray}
& {U_* \over n_e k_{\rm B} T_e} = \alpha_2 (3+\alpha_2)
\bigg[{2 \, \xi_c^{(2-\alpha_2)/2} \, \Gamma\left({3+\alpha_2 \over 2}
, \, \xi_c \right) \over \sqrt{\pi} \, (2-\alpha_2) (3+2\alpha_2)}
\nonumber \\
& \phantom{lotsofspaaace} + \, {2 \sqrt{\pi \xi_c} \, (3 + 2 \, \xi_c)
\, e^{-\xi_c} + \, 3 \pi \, {\rm Erfc}\left(\xi_c^{1/2}\right)
\over 2 \pi (\alpha_2^2 + 3 \alpha_2 - 10)}\bigg] \ ,
\label{neweq24}
\end{eqnarray}
They also show that the total
power $L_{\rm in}$ required to drive the acceleration of the nonthermal electrons as
\begin{equation}
L_{\rm in} = 8 V D_0 \, U_* \ ,
\label{neweq25}
\end{equation}
where $V$ is the volume of the acceleration region. Using reasonable estimates for the volume of the acceleration region and using values for the power $L_{\rm out}$ in the observed radio emission, Subramanian \& Becker[60] have concluded that the efficiency of the overall plasma emission process starting from electron acceleration and culminating in the observed noise storm emission is $\eta \equiv L_{\rm out}/L_{\rm in} < 10^{-6}$.

\vspace{.4cm}
\noindent
{\s 6. Discussion and conclusions:}
\\
\\
Apart from type 1/noise storm emission, the plasma emission process is thought to be operative in several types of meter wavelength emission such type II and type III emission (e.g., Melrose[37]). Accurate estimates of the efficiency of the overall plasma emission process are important, for instance, in deducing the shock strength from the intensity of observed type II emission. If the observed intensity can be well related to the energy in the nonthermal electrons, one can place reliable constraints on the parameters of the shock that produced the nonthermal electrons. The relationship of shocks in the heliosphere to the coronal mass ejections that presumably drive it is not very clear. This is of crucial practical importance, since such shocks are thought to be responsible for a variety of space weather-related effects in the near-earth space environment.

There are yet interesting aspects with regard to the role of electron acceleration in the plasma emission process that remain to be understood. One of them can be stated as follows: multifrequency observations of noise storms (\S~2.1) reveal that the observed intensity is usually an increasing function of frequency in the frequency range 30 -- approximately 100 MHz. However, at higher frequencies, (higher than $\sim$ 150 MHz) the observed intensity is a decreasing function of frequency. It is interesting to consider the implications of this apparent peak in the observed noise storm intensity at around 100--150 MHz. The observed frequency corresponds directly to height in the solar corona; the lower the frequency, the larger the height. The observed frequency can therefore be considered as a proxy for height. Do the multifrequency noise storm observations imply there is an optimum height for quasi-continuous electron acceleration in the solar corona? If so, why? The background density decreases with height in the solar corona, and with it so do the collisional losses. On the other hand, this also means that the density of the parent pool of electrons available for acceleration from the thermal tail also decreases with height. Do these competing effects yield an optimum height for electron acceleration? There is also a similar, fairly well accepted notion regarding electron acceleration in solar flares (Aschwanden \& Benz[66]); it states that there is a ``primary'' acceleration height in the solar corona corresponding to a plasma frequency of $\sim$ 500--600 MHz. They arrive at this conclusion from observations of bi-directional type III bursts, which lead them to infer that electrons are primarily accelerated at the 500--600 MHz layer. Could this also be because the 500--600 MHz layer is somehow an optimum place for electron acceleration under these conditions, and could it be for reasons similar to the peak in noise storm emission at around 100--150 MHz? These are some of the questions that remain to be answered in this interesting sub-field of plasma astrophysics.

\vspace*{0.8cm} 
\noindent
\vspace{.4cm}
\noindent
{\s References}
\\
{\fc
\noindent
1 Bastian, TS, Gary, DE, in {\it From Clark Lake to the Long Wavelength Array: Bill Erickson's Radio Science}, ASP Conference Series, Vol. 345, eds. Kassim, NE, Perez, MR, Junor, W, Henning, PA, Astron. Soc. Pacific (2005).

\noindent
2 Elgaroy EO, {\it Solar Noise Storms}, Pergamonn Press, Oxford (1977).

\noindent
3 {\it Solar and Space Weather Radiophysics - Current Status and Future Developments}, eds. Gary DE, Keller CU, Kluwer Academic Publishers, Dordrecht (2004).

\noindent
4 Hey JS, {\it Nature}, 157 (1946), 47.

\noindent
5 {\it Solar Radiophysics}, eds. McLean DJ, Labrum NR, Cambridge University
   Press, Cambridge (1985).

\noindent
6 Sundaram GAS, Subramanian KR, {\it Astrophys. J.}, 605 (2004), 948.

\noindent
7 Smerd SF, {\it Ann. of the IGY}, 34 (1964), 331.

\noindent
8 Wild JP, Smerd SF, Weiss AA, {\it Ann. Rev. Astron. Astrophys.}, 1 (1963), 291.

\noindent
9 Kundu MR, {\it Solar Radio Astronomy}, Interscience, NY (1965).

\noindent
10 Kruger A, {\it Introduction to Solar Radio Astronomy and Radio Physics}, Reidel, Dordecht (1979).

\noindent
11 Dulk GA, Nelson GJ, {\it Proc. Astron. Soc. Australia}, 2 (1973), 211.

\noindent
12 Kundu MR, Gopalswamy N, {\it Solar Phys.}, 129 (1990), 133.

\noindent
13 Thejappa G, Kundu MR,  {\it Solar Phys.}, 132 (1991), 155.

\noindent
14 Kerdraon A, Mercier C, {\it Astron. Astrophys.}, 127 (1983), 132.

\noindent
15 Malik RK, Mercier C, {\it Solar Phys.}, 165 (1996), 347.

\noindent
16 Willson RF, Kile JN, Rothberg B, {\it Solar Phys.}, 170 (1997), 299.

\noindent
17 Willson RF, {\it Solar Phys.}, 197 (2000), 399.

\noindent
18 Habbal SR, Ellman NE, Gonzalez R, {\it Astrophys. J.}, 342 (1989), 594.

\noindent
19 Habbal SR, Mossman A, Gonzalez R, Esser R, {\it J. Geophys. Res.}, 101 (1996), 19943.

\noindent
20 Mercier C, Subramanian P, Kerdraon A, Pick M, Ananthakrishnan S, Janardhan P, {\it Astron. Astrophys.}, 447 (2006), 1189.

\noindent
21 Bastian TS, Pick M, Kerdraon A, Maia D, Vourlidas A, {\it Astrophys. J.}, 558 (2001), L65.

\noindent
22 Willson RF, {\it Solar Phys.}, 211 (2002), 289.

\noindent
23 Willson RF, {\it Solar Phys.}, 227 (2005), 311.

\noindent
24 Bastian TS, {\it Astrophys. J.}, 426 (1994), 774.

\noindent
25 Zlobec P, Messerotti M, Dulk GA, Kucera T, {\it Solar Phys.}, 141 (1992), 165.

\noindent
26 Robinson RD, {\it Astrophys. J.}, 222 (1978), 696.

\noindent
27 Kerdraon A, {\it Astron. Astrophys.}, 71 (1979), 266.

\noindent
28 Klein K.-L, in {\it Second Advances in Solar Physics Euroconference: Three-Dimensional Structure of Solar Active Regions}, ASP Conf. Series Vol. 155, eds. Alissandrakis CE, Schmieder B, Astron. Soc. Pacific (1998).

\noindent
29 Raulin JP, Klein, K.-L, {\it Astron. Astrophys.}, 281 (1994), 536.

\noindent
30 Kerdraon A, Pick M, Trottet G, Sawyer C, Illing R, Wagner W, House L, {\it Astrophys. J.}, 265 (1983), L19.

\noindent
31 Bentley RD, Klein K.-L, van Driel-Gesztelyi L, Demoulin P, Trottet G, Tassetto P, Marty G, {\it Solar Phys.}, 193 (2000), 227.

\noindent
32 Priest E, Forbes T, {\it Magnetic Reconnection: MHD Theory and Applications}, Cambridge Univ. Press, Cambridge (2000).

\noindent
33 Biskamp D, {\it Magnetic Reconnection in Plasmas}, Cambridge Univ. Press, Cambridge (2000).

\noindent
34 Spicer DS, Benz AO, Huba JD, {\it Astron. Astrophys.}, 105 (1982), 221.

\noindent
35 Mikhailovskii AB, {\it Theory of Plasma Instabilities}, Consultant Bureau, New York (1974).

\noindent
36 Melrose DB, {\it Solar Phys.}, 111 (1987), 89.

\noindent
37 Melrose DB, {\it Solar Phys.}, 43 (1975), 211.

\noindent
38 Melrose DB, {\it Solar Phys.}, 67 (1980), 357.

\noindent
39 Wentzel DG, {\it Solar Phys.}, 103 (1986), 141.

\noindent
40 Krucker S., Benz AO, Aschwanden MJ, Bastian TS, {\it Solar Phys.}, 160 (1995), 151.

\noindent
41 Mercier C, Trottet G, {\it Astrophys. J.}, 474 (1997), L65.

\noindent
42 Thejappa G, {\it Solar Phys.}, 132 (1991), 173.

\noindent
43 Aschwanden MJ, {\it Physics of the Solar Corona}, Springer, Berlin (2004).

\noindent
44 O'C Drury L, {\it Rep. Prog. Phys.}, 46 (1983), 973.

\noindent
45 Quenby JJ, Meli A, in {\it Particle acceleration in astrophysical plasmas: geospace and beyond}, Geophysical monograph, vol. 156, American Geophysical Union, Washington, DC (2005).

\noindent
46 Malkov MA, O'C Drury L, {\it Rep. Prog. Phys.}, 64 (2001), 429.

\noindent
47 Jones FC, Ellison DC, {\it Space Sci. Rev.}, 58 (1991), 259.

\noindent
48 Onofri M, Isliker H, Vlahos L, {\it Phys. Rev. Lett.}, 96 (2006), 151102.

\noindent
49 Turkmani R, Cargill PJ, Galsgaard K, Vlahos L, Isliker H, {\it Astron. Astrophys.}, 449 (2006), 749.

\noindent
50 Arzner K, Knaepen B, Carati D, Denewet N, Vlahos L, {\it Astrophys J.}, 637 (2006), 322.

\noindent
51 Yamada M, Ji H, Kulsrud R, Kuritsyn A, Ren Y, Gerhardt S, Breslau J, in {\it Magnetic fields in the Universe: From Laboratory and Stars to Primordial Structures}, AIP Conf. Proc. vol. 784, American Institute of Physics, New York (2005).

\noindent
52 Sturrock PA, {\it Plasma Physics}, Cambridge Univ. Press, Cambridge (1994).

\noindent
53 Montgomery DC, Tidman DA, {\it Plasma Kinetic Theory}, McGraw-Hill, New York (1964).

\noindent
54 Subramanian P, Becker PA, Kazanas D, {\it Astrophys. J.}, 523 (1999), 203.

\noindent
55 Longair MS, {\it High Energy Astrophysics: vol 2}, Cambridge Univ. Press, Cambridge (1994).

\noindent
56 Becker PA, Le T, Dermer CD, {\it Astrophys. J.}, 647 (2006), 539.

\noindent
57 Becker PA, {\it Astrophys. J.}, 397 (1992), 88.

\noindent
58 Schlikeiser R, {\it Cosmic Ray Astrophysics}, Springer, Berlin (2002).

\noindent
59 Subramanian P, Becker PA, {\it Solar Phys.}, 225 (2004), 91.

\noindent
60 Subramanian P, Becker PA, {\it Solar Phys.}, 237 (2006), 185.

\noindent
61 Smith DF, {\it Astrophys. J.}, 212 (1977), 891.

\noindent
62 Ptuskin VS, {\it Soviet Astron. Lett.}, 14 (1988), 255.

\noindent
63 Chandran BDG, Maron JL, {\it Astrophys. J.}, 603 (2003), 23.

\noindent
64 Miller JA, LaRosa TN, Moore RL, {\it Astrophys. J.}, 461 (1996), 445.

\noindent
65 Luo QY, Wei FS, Feng XS, {\it Astrophys. J.}, 584 (2003), 497.

\noindent
66 Aschwanden MJ, Benz AO, {\it Astrophys. J.}, 480 (1997), 825.

\end{document}